\documentclass[graybox]{svmult}

\usepackage{helvet}         
\usepackage{courier}        
\usepackage{type1cm}        
\usepackage{makeidx}         
\usepackage{graphicx}        
\usepackage{multicol}        
\usepackage[bottom]{footmisc}

\usepackage{amsmath, amssymb}

\usepackage{listings}
\usepackage{lstautogobble}
\usepackage{lstlinebgrd}

\usepackage[scaled]{inconsolata}
\usepackage{xcolor}
\definecolor{keywordcolor}{rgb}{0.7, 0.1, 0.1}   
\definecolor{tacticcolor}{rgb}{0.0, 0.1, 0.6}    
\definecolor{commentcolor}{rgb}{0.4, 0.4, 0.4}   
\definecolor{symbolcolor}{rgb}{0.0, 0.1, 0.6}    
\definecolor{sortcolor}{rgb}{0.1, 0.5, 0.1}      
\definecolor{attributecolor}{rgb}{0.7, 0.1, 0.1} 

\begin{document}

\title*{The mechanization of science illustrated by the Lean formalization of the multi-graded Proj construction}
\titlerunning{Lean and Proj} 

\author{Arnaud Mayeux \and Jujian Zhang}

\institute{Arnaud Mayeux \at Einstein Institute of Mathematics, The Hebrew University of Jerusalem\\ \email{arnaud.mayeux@mail.huji.ac.il}
\and Jujian Zhang \at Department of Mathematics, Imperial College London\\ \email{jujian.zhang19@imperial.ac.uk}}

\maketitle

Efforts to mechanize aspects of scientific reasoning have been intertwined with the development of science from its earliest days. In a 1961 paper \textquotedblleft The Mechanization of Science\textquotedblright Hamming discusses several steps:
 
 \begin{small}
 
\begin{enumerate}
\item[C1] \textit{\textquotedblleft Whenever we have a long, difficult piece of algebra, and we have them more and more often these days, we could at least get the machine to check that the algebra was right before we went on and built further stages of derivation on top. Some people are working on such programs for algebra checking right now.\textquotedblright}

\item[C2] \textit{\textquotedblleft Now I leave the region of known processes and enter the land of speculation. We can, I believe, reasonably expect that an algebra checking routine would not be around very long before someone would adapt the methods of heuristics that are presently being developed to the problem of doing algebra in a more creative way. The machine could supply several steps at a time, and be given only a guiding thread of a proof. The more successful the heuristics, the fewer steps we would have to supply.\textquotedblright}

\item[C3] \textit{\textquotedblleft How far we will go depends on the success of the people working in the fields of theorem proving, pattern recognition, and general heuristics. What has been achieved has been remarkable, but I certainly feel that we are a long way from being able to give the machine an important theorem for which we do not know any proof, and expect to get one from it.\textquotedblright}

\item[C4] \textit{\textquotedblleft Suppose, for the moment, we did have such a powerful routine that when we gave the machine the Riemann Hypothesis it finally gave out some 500 pages of closely printed detail purporting to be a proof. Suppose further, that we have an independent routine called a 'theorem-proof checker.' Such a routine would be much easier to write than the first one. And suppose the alleged proof passed the second test. Would that constitute a proof of the theorem? [...] I tend to feel that the pragmatic result would be that some people would build on top of it, and hence use it as a proof, thus for them it would be a proof whether any human ever followed and understood it or not.\textquotedblright}

\item[C5] \textit{\textquotedblleft I have been talking about mathematics. While the details will be different in other sciences, I feel that much can be done to mechanize them. I can imagine a routine which can take a conjecture in modern physics and follow out a great many of the logical consequences, and can make a number of checks against known results. Thus the physicist in some areas could spend much of his time formulating new models, and letting the machine give the consequences.\textquotedblright}
\end{enumerate}
 \end{small}
These citations are extracted directly from \cite{Ha61}. Comparable thoughts can be found in many other antecedent and subsequent texts. Further consequences and visions are abundant too.
Formalization theorem provers like Lean (cf. \cite{dMal15, dMU21, Av24, Bu24}) fulfill the condition of C1. Lean is enough easy to use to be called a \textquotedblleft theorem-proof checker\textquotedblright (cf. C4).  In this note, we discuss the Lean formalization of the multi-graded Proj construction \cite{MZ25}. The multi-graded Proj construction sits at the crossroads of algebraic geometry and Lie theory \cite{BS03,MR24}, both fields actively used in theoretical physics.
The Lean4 code of the formalization of the multi-graded Proj construction is around 8000 lines, for 400 lines of \LaTeX.
This uses much of the material on the formalisation of scheme theory available in mathlib \cite{Z23, Bual22, WZ22, Mathlib}.
It would be useful to see advances in AI-automated formalisation and a more visual interface for formalisation (e.g. \LaTeX-friendly).
As of Spring 2025, Copilot (Lean 2025 AI) remained insufficiently effective for assisting with the formalisation of algebraic geometry. Copilot was not able to fill by itself the proof of the fact that $0 \times 1 = 0$ in a certain non-trivial graded ring, because the simple calculation involves proofs about degree as well. In relation with C2 and C3 above, the present work \cite{MZ25}, as new formalised data, will contribute to the training of these already powerful tools.
We now introduce the algebraic part of this note. 
Inspired by projective geometry, Grothendieck defined the Proj scheme of an $\mathbf{N}$-graded ring $A$ \cite{Gr61}. This construction is used and presented in thousands of publications. 
Beyond the definition of Proj, most treatments also study quasi-coherent sheaves and twisting sheaves on Proj and use Proj to define blowups.  
Grothendieck Proj construction was formalised in Lean by Zhang \cite{Z23}. 
So it appears that this Proj construction is already highly studied.  
However, one can ask the following question. Why Grothendieck assumed that $A$ is $\mathbf{N}$-graded instead of graded by a general commutative monoid or group? 
Our answer is: there is no reason from the point of view of scheme theory. In fact, Brenner-Schroer defined the Proj of a ring graded by an arbitrary finitely generated abelian group \cite{BS03}.
To our knowledge, as of 2025, textbooks have never mentioned the work of Brenner and Schroer.
 Informal versions of the multi-graded Proj construction appear implicitly or
explicitly (but without reference to Brenner-Schroer) in various constructions in mathematics (cf. \cite{MR24}).
The first part of this note explains Brenner-Schroer construction and new results. The second part discusses the formalization. This note is based on two similar talks delivered at different places around the same time, one on 19 June 2025 at the seminar {\small \textit{Formalisation of mathematics with interactive theorem provers, University of Cambridge}} and one at the conference {\small \textit{XVI. Lie theory and its applications in physics, 16 - 22 June 2025, Varna}}.
\begin{flushleft}
\textbf{{1. Definition of Proj and Behaviors}}
\end{flushleft}
This section is based on \cite{MR24}. The reference \cite{BS03} introduced the main definition, and \cite{MR24} subsequently developed the theory further, providing additional reformulations and results.
If a given commutative ring $A$ is $M$-graded for some abelian group $M$, we will say that a submonoid $S$ of $A$ is homogeneous if every element of $S$ is homogeneous. 
In this situation, the localization $A_S$ of $A$ with respect to $S$ is canonically $M$-graded.
Given a graded $A$-module $Q$, one can also consider the localized module $Q_S$, which has a natural structure of a graded $A_S$-module.
Given a homogeneous submonoid $S \subset A$, we will denote by $\underline{S}$ the homogeneous submonoid consisting of homogeneous divisors of elements in $S$. Note that we have a canonical isomorphism of graded rings
 $A_{ \underline{S}} \cong A_{S}.
$
If $Q$ is a graded $A$-module we also have an identification
$
Q_{\underline{S}} \cong Q_S
$
compatible with the actions of $A_{ \underline{S}} = A_{S}$.
Consider an abelian group $M$, and a commutative $M$-graded ring $A$. 

\vspace{0.25\baselineskip}
\noindent \textbf{Definition:}
 Let $S$ be a homogeneous subset of $A$. We denote by $\deg(S)$ the subset of $M$ defined as 
$\deg(S):= \{ m \in M : \exists s \in S , s  \in A_m \}. $

Note that $\deg(\{0\})=M$.
More generally, if $S$ is a homogeneous submonoid of $A$, then $\deg(S)$ is a submonoid of $M$, also denoted $M[S\rangle$.

\vspace{0.25\baselineskip}
\noindent \textbf{{Definition:}}
Let $S$ be a homogeneous submonoid of $A$. 
 We put $M[S]=M[S \rangle^{\mathrm{gp}}= M[S \rangle- M[S \rangle$, the subgroup of $M$ generated by $\deg(S)$.

\vspace{0.25\baselineskip}
\noindent\textbf{{Definition:}}
 A homogeneous submonoid $S$ of $A$ is called \emph{$M$-relevant} (or just \emph{relevant} if $M$ is clear from the context) if for any $m$ in $M$ there exists $n \in \mathbf{Z}_{> 0}$ such that $nm$ belongs to $M[\underline{S}]$, i.e. if $M/(M[\underline{S}])$ is a torsion abelian group.
 A family $\{a_i : i \in I\}$ of homogeneous elements in $A$ is called $M$-relevant if the submonoid generated by the $a_i$'s is relevant.
 A homogeneous element $a \in A$ is called $M$-relevant if the family $\{a\}$ is $M$-relevant.

Let $S$ be a homogeneous submonoid of $A$.

\vspace{0.25\baselineskip}
\noindent
\textbf{\textbf{Definition:}}
The degree-$0$ part $(A_S)_0$ of the localization $A_{S}$ is denoted $A_{(S)}$ and is called the \emph{potion} of $A$ with respect to $S$. 
If $Q$ is a graded $A$-module, we will also denote by $Q_{(S)}$ the degree-$0$ part $(Q_S)_0$ of $Q_S$; it admits a canonical structure of an $A_{(S)}$-module.

We have canonical identifications
$A_{(\underline{S})} \cong A_{(S)}$ and $Q_{(\underline{S})} \cong Q_{(S)}$.
If  $\{a_i : i \in I \}$ is a family of homogeneous elements of $A$, we will denote by
$A_{(\{a_i : i \in I \})}$
the potion associated with the submonoid of $A$ generated by $\{a_i : i \in I \}$; in case $\#I=1$ we will write $A_{(a)}$ for $A_{(\{a\})}$.
If $S$ and $T$ are submonoids of $A$, we will denote by $ST$ the submonoid of $A$ generated by $S \cup T$, i.e. $ST = \{st : s \in S, \, t \in T \}$. Of course, $ST$ is homogeneous if $S$ and $T$ are.
The following is the key result that makes the Proj construction work.

\vspace{0.25\baselineskip}
\noindent
\textbf{{Proposition (Magic of potions):}}
Let $S$ and $T$ be homogeneous submonoids of $A$. 
\begin{enumerate}
\item 
\label{it:magic-1}
We have a canonical homomorphism of potion rings 
$A_{(S )} \to  A_{(ST )}$.
\item 
\label{it:magic-2}
Assume that $S$ is relevant. Fix a subset $T' \subset T$ which generates $T$ as a submonoid of $(A,\times)$ and, for any $t $ in $T'$, fix $n_t \in \mathbf{N}_{>0}$ and $s_t, s_t' \in \underline{S}$ such that $\deg(t^{n_t}) = \deg(s_t)-\deg(s_t')$.
Then $\frac{t^{n_t} s_t'} { s_t}$ belongs to $A_{(\underline{S} )}\cong A_{(S)}$. 
 Moreover we have a canonical isomorphism of $A_{(S )}$-algebras between $ A_{(ST )}$ and the localization of $A_{(S )}$ with respect to the submonoid of $A_{(S)}$ generated by $\{\frac{t^{n_t} s_t'} { s_t} : t \in T' \}$.
\item 
\label{it:magic-3}
Assume that $S$ is relevant and that $T$ is finitely generated as submonoid of $(A,\times)$. The morphism of schemes 
$
\mathrm{Spec} (A_{(ST )}) \to \mathrm{Spec} (A_{(S)})
$
induced by the ring homomorphism in (1) is an open immersion of schemes.
\item 
\label{it:magic-4}
Let $f_1, \ldots, f_n \in A$ be nonzero relevant homogeneous elements of the same degree.
Then we have a canonical open immersion
\[
\mathrm{Spec}(A_{ (f_1 + \cdots + f_n) }) \to \mathrm{Spec} (A_{(f_1)}) \cup \cdots \cup \mathrm{Spec} (A_{(f_n)})
\]
where the right-hand side is defined as the glueing of the affine schemes $\mathrm{Spec} (A_{(f_i)})$ along the open subschemes $\mathrm{Spec} (A_{(f_i \cdot f_j)}) \subset \mathrm{Spec} (A_{(f_i)})$.
\end{enumerate}

From now on we assume that $M$ is \emph{finitely generated}, and fix a commutative $M$-graded ring $A$.
We denote by $\mathcal{F}_A$ the set of all relevant homogeneous submonoids of $A$ which are finitely generated as submonoids of $(A,\times)$.

\vspace{0.25\baselineskip}
\noindent
\textbf{{Construction (Proj as glueing potions)}}
Let
$\mathcal{F} \subset \mathcal{F}_A$ be a subset.
 For each $S \in \mathcal{F}$, let $D_{\dagger}(S)$ be the spectrum of the potion $A_{(S)}$. 
 If $S,T \in \mathcal{F}$, the affine scheme $D_{\dagger}(ST)$ identifies canonically with an open subscheme of $D_{\dagger}(S)$. For each $S,T \in \mathcal{F}$, we have equalities
$
 D_{\dagger}({S S}) = D_{\dagger}(S) $ and $ D_{\dagger}({ST})=D_{\dagger}({TS}).
$
 Moreover, for each triple $S,T,U \in \mathcal{F}$, we have
$
 D_{\dagger}({ST} )\cap D_{\dagger}({SU}) = D_{\dagger}({TS}) \cap D_{\dagger}({TU}).
$
Now, by glueing, from these data we obtain a scheme $\mathrm{Proj}^M_{\mathcal{F} } (A)$ and, for each $S \in \mathcal{F}$, an open immersion $\varphi_S : D_{\dagger}(S) \to \mathrm{Proj}^M_{\mathcal{F}} (A)$, such that
$
\mathrm{Proj}^M_{\mathcal{F} } (A) = \bigcup_{S \in \mathcal{F} } \varphi_S( D_{\dagger}(S)).
$
In practice, we will often identify $D_{\dagger}(S) $ and $\varphi_S (D_{\dagger}(S))$. 
In the case when $\mathcal{F} = \mathcal{F}_A$,
 the scheme $\mathrm{Proj}^M_{\mathcal{F}_A} (A)$ will be denoted $\mathrm{Proj}^M(A)$, or just $\mathrm{Proj}( A)$ when $M$ is clear from the context.

\vspace{0.25\baselineskip}
\noindent
\textbf{{Proposition:}}
The scheme $\mathrm{Proj}^M(A)$ is quasi-separated.

\vspace{0.25\baselineskip}
\noindent
{\textbf{Proposition (Functoriality of Proj):}} 
Let $\Psi:A \to B $ be a homomorphism of $M$-graded rings. 
For any $\mathcal{F} \subset \mathcal{F}_A$ 
we have a canonical morphism of schemes $\mathrm{Proj}^M_{\Psi (\mathcal{F})} (B) \to \mathrm{Proj}^M_{\mathcal{F} } (A)$.

\vspace{0.25\baselineskip}
\noindent
\textbf{{Proposition:}}
Assume that $\Psi : A \to B$ is surjective. Then we have $\mathrm{Proj}^M_{\Psi (\mathcal{F}_A)} (B) = \mathrm{Proj}^M (B)$, and the canonical morphism
 $\mathrm{Proj}^M(B) \to \mathrm{Proj}^M(A)$
is a closed immersion.

\vspace{0.25\baselineskip}
\noindent
\textbf{{Proposition:}}
 Let $M$ and $M'$ be two finitely generated abelian groups. Let $R$ be a commutative ring and let $A$ (resp.~$A'$) be a commutative $M$-graded (resp.~$M'$-graded) $R$-algebra. 
 Then for the natural $(M \times M')$-grading on $A \otimes_R A'$, we have a canonical isomorphism
 \[
 \mathrm{Proj} ^{M \times M'} (A \otimes_R A') \cong \mathrm{Proj} ^M (A) \times_{\mathrm{Spec}(R)} \mathrm{Proj} ^{M'} (A'). 
 \]

Let $Q$ be an $M$-graded $A$-module. For any homogeneous submonoid $S \subset A$, we have considered the $A_{(S)}$-module $Q_{(S)}$. 

\vspace{0.25\baselineskip}
\noindent
\textbf{{Definition:}}
There exists 
a unique quasi-coherent $\mathcal{O}_{\mathrm{Proj}^M(A)}$-module $\widetilde{Q}$ such that
$\Gamma \bigl( D_\dag(S)  , \widetilde{Q} \bigr) = Q_{(S)}$
for every $S \in \mathcal{F}_A$.

An $M$-graded $A$-module $Q$ will be called \emph{negligible} if $\widetilde{Q}=0$. 
If $Q$ is a graded $A$-module and $\alpha \in M$, we will denote by $Q(\alpha)$ the $M$-graded $A$-module which coincides with $Q$ as an $A$-module, but with the $M$-grading defined by $(Q(\alpha))_\beta = Q_{\alpha+\beta}$ for $\beta \in M$.

\vspace{0.25\baselineskip}
\noindent
{\textbf{Definition (Twisting sheaves)}}
Let $\alpha \in M$.
The quasi-coherent sheaf $\widetilde{A(\alpha)}$ on $\mathrm{Proj}^M(A)$ is 
denoted $\mathcal{O}_{\mathrm{Proj}^M(A)} (\alpha)$.
If $\mathcal{Q}$ is a sheaf of $\mathcal{O}_{\mathrm{Proj}^M(A)}$-modules, we set $\mathcal{Q} (\alpha) = \mathcal{O}_{\mathrm{Proj}^M(A)} (\alpha ) \otimes_{\mathcal{O}_{\mathrm{Proj}^M(A)}} \mathcal{Q}$.

\vspace{0.25\baselineskip}
\noindent
\textbf{{Proposition:}}
Assume that $A$ is a noetherian ring.
For any $\alpha \in M$, the quasi-coherent sheaf $\mathcal{O}_{\mathrm{Proj}^M(A)} (\alpha)$ is coherent.
If $Q$ is a finitely generated $M$-graded $A$-module, then $\widetilde{Q}$ is coherent.

A family $S \in \mathcal{F}_A$ will be called \emph{maximally relevant} if $M[\underline{S}]=M$.   We will denote by $\mathcal{F}_A^{\mathrm{m}} \subset \mathcal{F}_A$ the subset consisting of maximally relevant families. We now discuss some consequences of the condition:
$
\mathrm{Proj}^M(A) = \bigcup_{S \in \mathcal{F}_A^{\mathrm{m}}} D_\dag(S).$

\vspace{0.25\baselineskip}
\noindent
\textbf{{Proposition:}}
Assume that the above condition is satisfied. For any $ \alpha \in M$, the quasi-coherent $\mathcal{O}_{\mathrm{Proj}^M(A)}$-module $\mathcal{O}_{\mathrm{Proj}^M(A)} (\alpha)$ is an invertible sheaf.

\vspace{0.25\baselineskip}
\noindent
\textbf{{Proposition:}}
Assume that the above condition is satisfied, and moreover that $\mathrm{Proj}^M(A)$ is quasi-compact. Then the functor
\[
\mathsf{L} : \mathrm{Mod}^M(A) / \mathrm{Mod}^M(A)_{\mathrm{neg}} \to \mathrm{QCoh}(\mathrm{Proj}^M(A))
\]
is an equivalence of categories.

\vspace{0.25\baselineskip}
\noindent
\textbf{Example (Flag variety as Proj)}
 Let $k$ be an algebraically closed field and let $G$ be
a connected reductive group scheme over $k.$ Let $T$ be a maximal split torus
and $B$ be a Borel subgroup such that $B = T N$ where $N$ is the unipotent
radical. Then $G/N$ is quasi-affine and the ring $A := \Gamma (G/N, \mathcal{O}_{G/N} )$ is canonically $X^*(T)$-graded. We have a schematically dominant open immersion
$G/N \to \mathrm{Spec}(A)$. 
The flag variety $G/B$ identifies with $\mathrm{Proj}^{X^*(T)} (A)$.
Consider a finite-dimensional $G$-module $\widetilde{V}$ and a $B$-stable subspace $V \subset \widetilde{V}$. 
We can then consider the induced scheme
$
 G \times^{B} V,
$
i.e.~the quotient of the product $G \times V$ by the (free) action of $B$ defined by $b \cdot (g,x) = (gb^{-1}, b \cdot x)$.  It is a vector bundle over $G/B$.
There is a canonical way to see such schemes as Proj too.
An example is when $\widetilde{V}$ is the Lie algebra of $G$ and $V$ is the Lie algebra of the unipotent radical of $B$. Then $G \times^{B} V$ is the Springer resolution.

\begin{flushleft}
\textbf{2. Formalisation}
\end{flushleft}
Compared to sets and subsets, quotient type is a fundamental concept in the type system of Lean4. 
Therefore, it is more natural and ergonomic to define the degree-$0$ part of a graded ring as a quotient type 
instead of a subset of the graded ring. For an $\iota$-graded ring $A \cong \bigoplus_{i : \iota} \mathcal{A}_i$ and a submonoid $x \le A$, 
we define the degree-$0$ part of the localization $A_x$ to be the equivalence classes of triples $(i, n, d)$ where 
$i : \tau$ and both $n$ and $d$ are elements of $\mathcal{A}_i$ under the equivalence relation 
$(i, n, d) \sim (i', n', d')$ if and only if the equality $\frac{n}{d} = \frac{n'}{d'}$ in the localized ring $A_x$:

\begin{lstlisting}
structure NumDenSameDeg where
  deg : ι
  (num den : 𝒜 deg)
  den_mem : (den : A) ∈ x

def NumDenSameDeg.embedding (p : NumDenSameDeg 𝒜 x) :=
  Localization.mk p.num ⟨p.den, p.den_mem⟩

def HomogeneousLocalization : Type _ :=
  Quotient (Setoid.ker <| NumDenSameDeg.embedding 𝒜 x)
\end{lstlisting}

The advantage of the quotient approach is two-fold:
\begin{itemize}
  \item The numerator $n$, the denominator $d$, the degree $i$ and the fact that $n, d \in \mathcal{A}_i$ and $d \in x$ and  are one application of axiom of choice away.
  Compared to the set-theoretic approach $A_{(x)} = \left\{f\middle|\exists i \in \iota, n \in \mathcal{A}_i \cap x, d \in \mathcal{A}_i, f = \frac{n}{d}\right\}$, the extraction of the data needs repeated application of AC.
  \item The quotient type comes with a universal property already making defining functions out of $A_{(x)}$ convenient.
\end{itemize}

The practice of formalisation often forces a clearer statement.
In the second part of the magic of potions, we can define the following data type:

\begin{lstlisting}
structure PotionGen where
  (index : Type*)
  (elem : index → A)
  (elem_mem : ∀ t, elem t ∈ T)
  (gen : Submonoid.closure (Set.range elem) = T.toSubmonoid)
  (n : index → ℕ+)
  (s s' : index → A)
  (s_mem_bar : ∀ t, s t ∈ S.bar)
  (s'_mem_bar : ∀ t, s' t ∈ S.bar)
  (i i' : index → ι)
  (t_deg : ∀ t : index, (elem t : A)^(n t : ℕ) ∈ 𝒜 (i t - i' t))
  (s_deg : ∀ t, s t ∈ 𝒜 (i t))
  (s'_deg : ∀ t, s' t ∈ 𝒜 (i' t))

def PotionGen.genSubmonoid (T' : PotionGen S T) : Submonoid S.Potion :=
  Submonoid.closure
    {x | ∃ (t : T'.index), x =
      S.equivBarPotion.symm (.mk
        { deg := T'.i t,
          num := ⟨(T'.elem t) ^ (T'.n t : ℕ) * T'.s' t,
            by simpa using SetLike.mul_mem_graded (T'.t_deg t) (T'.s'_deg t)⟩,
          den := ⟨T'.s t, T'.s_deg t⟩,
          den_mem := T'.s_mem_bar t }) }
\end{lstlisting}

Therefore, the statement of the second part of the magic of potions can be stated as:
for every finitely generated, relevant homogeneous submonoid $S, T$ of $A$ and any \lstinline|PotionGen S T| $T'$, we have an isomorphism between
the potion $A_{(ST)}$ and the localized ring of $A_S$ at genSubmonoid $T'$:

\begin{lstlisting}
def localizationRingEquivPotion (T' : PotionGen S T) :
    Localization T'.genSubmonoid ≃+* (S * T).Potion := ...
\end{lstlisting}
Of course, in a pen-and-paper proof, one could define auxiliary structures to make the same effect, 
but the absence of such structures does not put the burden to the author, therefore, the author is 
less motivated to define such structures. However, in formalisation, 
the burden is put on the authors --- without such structures, the proof is not only less readable to the reader, but much harder to write for the author as well.

On the other hand, the clarity and rigour brought by formalisation \textit{a priori} requires extra 
care from the formaliser --- steps seen as unnecessary on pen-and-paper suddenly need to 
be explained to computers. One can admit that this is because type theory is not the same as set theory and 
one should not expect verbatim practices and there are many places where intermediate steps are indeed missing.
A noticeable difference is the notion of equality --- consider the following example, if $S$ and $T$ are two equal homogeneous submonoids of $A$, 
we should be able to treat $A_{(S)}$ and $A_{(T)}$ as the same object in some sense. 
Prior to formalisation, we thought $A_{(S)}$ and $A_{(T)}$ are equal, but in set theory, $A_{(S)}$ and $A_{(T)}$ are subrings of $A_S$ and $A_T$ respectively,  
therefore, the literal equality $A_{(S)} = A_{(T)}$ is simply not true. 
In type theory, though we can prove the equality $A_{(S)} = A_{(T)}$ holds, the equality provides less useful functions because of the excessive need of \lstinline|rewrite| of the equality $S = T$.
Therefore, $A_{(S)}$ and $A_{(T)}$ are the same because the following isomorphism, note that both directions of the isomorphism are constructed
using the universal property of quotient types and not directly from the equality $S = T$:

\begin{lstlisting}
def potionEquiv {S T : HomogeneousSubmonoid 𝒜} (eq : S = T) : 
    S.Potion ≃+* T.Potion := 
  RingEquiv.ofHomInv
    (HomogeneousLocalization.map _ _ (RingHom.id _) ... : 
      S.Potion →+* T.Potion)
    (HomogeneousLocalization.map _ _ (RingHom.id _) ... : 
      T.Potion →+* S.Potion)
    ...
\end{lstlisting}

The requirement of spelling out ``trivial'' isomorphisms is not only an artifact of type theory, 
it is needed for rigour: consider the isomorphism between Cartesian product $A \times B \cong B \times A$, in case of $A = B$,
without more details $A \times A \cong A \times A$ is ambiguous.


\begin{thebibliography}{99.}


\bibitem[Av24]{Av24} J.\,Avigad, {\it Automated reasoning for mathematics,} in {\it Automated reasoning. Part I}, 3--20, Lecture Notes in Comput. Sci. Lecture Notes in Artificial Intelligence, 14739 , Springer, Cham, 2024.

\bibitem[BS03]{BS03} H.\,Brenner, S.\,Schroer, {\it Ample families, multihomogeneous spectra, and algebraization of formal schemes} Pacific J. Math. 208 (2003), no. 2, 209–230.

\bibitem[Bual22]{Bual22} K.\,Buzzard, C.\,Hughes, K. Lau, A\,
              Livingston, R. F. Mir, and S.\,Morrison, {\it Schemes in Lean,} Exp. Math. {\bf 31} (2022), no.~2, 355--363.
                 
              
              
\bibitem[Bu24]{Bu24} K.\,Buzzard, {\it Mathematical reasoning and the computer,} Bull. Amer. Math. Soc. (N.S.) {\bf 61} (2024), no.~2, 211--224.  



\bibitem[Gr61]{Gr61} A.\,Grothendieck, {\it EGA : II. Etude globale elementaire de quelques classes de morphismes} Publications mathématiques de l'I.H.E.S., tome 8 (1961), p. 5-222

\bibitem[Ha61]{Ha61} R.\,Hamming, {\it The mechanization of science} Proceedings of the 1961 16th ACM national meeting.     Association for Computing Machinery, New York, United States

\bibitem[Mathlib]{Mathlib} Lean Community, {\it Mathlib}, https://leanprover-community.github.io/mathlib-overview.html




\bibitem[MR24]{MR24}A. Mayeux and S. Riche, \textit{\it On multi-graded Proj schemes}, to appear in Publ. Res. Inst. Math. Sci. 



\bibitem[MZ25]{MZ25}A. Mayeux and J. Zhang, \textit{\it Multi-graded Proj construction in Lean4}, https://github.com/ProjConstruction/Proj (2025)

\bibitem[dMU21]{dMU21}
L.\,de\,Moura and S.\,Ullrich, {\it The Lean 4 theorem prover and programming language,} in {\it Automated deduction---CADE 28}, 625--635, Lecture Notes in Comput. Sci., 12699, Springer, Cham, 2021.

\bibitem[dMal15]{dMal15}
L.\,de\,Moura, S.\,Kong, J.\,Avigad, F.\,van\,Doorn and J.\,von\,Raumer, {\it The lean theorem prover (system description)}, in {\it Automated deduction---CADE 25}, 378--388, Lecture Notes in Comput. Sci. Lecture Notes in Artificial Intelligence, 9195 , 2015 Springer, Cham.

\bibitem[Z23]{Z23}J. Zhang, \textit{Formalising the Proj construction in Lean}, in {\it 14th International Conference on Interactive Theorem Proving}, Art. No. 35, LIPIcs. Leibniz Int. Proc. Inform., 268 (2023).

\bibitem[WZ22]{WZ22}
E.\,Wieser and J.\,Zhang, {\it Graded rings in Lean's dependent type theory,} in {\it Intelligent computer mathematics}, 122--137, Lecture Notes in Comput. Sci. Lecture Notes in Artificial Intelligence, 13467 , Springer, Cham, 2022.

\end{thebibliography}
\end{document}